\documentclass[12pt]{article}
\usepackage{array, booktabs}
\usepackage{bm}
\usepackage{graphicx}
\usepackage{caption}
\usepackage{amsmath,amssymb,wasysym}
\usepackage{algorithm}
\usepackage{multirow}
\usepackage{algpseudocode}
\usepackage{lineno}  
\usepackage{setspace}  

\usepackage[utf8]{inputenc}
\usepackage{xcolor}
\usepackage{hyperref}
\usepackage{cleveref}
\usepackage[margin=1 in]{geometry}

\usepackage[utf8]{inputenc}
\usepackage[T1]{fontenc}
\usepackage{helvet}

\usepackage[affil-it]{authblk}
\usepackage{lineno}  
\usepackage{setspace}  

\makeatletter
\newcommand*{\centerfloat}{%
  \parindent \z@
  \leftskip \z@ \@plus 1fil \@minus \textwidth
  \rightskip\leftskip
  \parfillskip \z@skip}
\makeatother

\DeclareMathOperator*{\argmax}{argmax}
\usepackage{xcolor}

\usepackage[
    backend=biber,natbib,
    style=nature,
]{biblatex}
\begin{document}

\title{Self-reorganization and Information Transfer  \\ in Massive Schools of Fish}
\author{Haotian Hang$^1$, Chenchen Huang$^{1}$, Alex Barnett$^2$, Eva Kanso$^{1,3}$\footnote{Corresponding author: Kanso@usc.edu}
\\
\footnotesize{
    $^1$Department of Aerospace and Mechanical Engineering,  University of Southern California,  Los Angeles, CA 90089\\%
    $^2$Center for Computational Mathematics, Flatiron Institute, New York City, NY 10010\\
    $^3$Department of Physics and Astronomy,  University of Southern California, Los Angeles, CA 90089\\
    }
}
\date{%
    \footnotesize{\today}
}


\date{\today}


\maketitle

The remarkable cohesion and coordination observed in moving animal groups and their collective responsiveness to threats are thought to be mediated by scale-free correlations, where changes in the behavior of one animal influence others in the group, regardless of the distance between them. But are these features independent of group size? Here, we investigate group cohesiveness and collective responsiveness in computational models of massive schools of fish of up to 50,000 individuals. We show that as the number of swimmers increases, flow interactions destabilize the school, creating clusters that constantly fragment, disperse, and regroup, similar to their biological counterparts. We calculate the spatial correlation and speed of information propagation in these dynamic clusters. Spatial correlations in cohesive and polarized clusters are indeed scale free, much like in natural animal groups, but fragmentation events are preceded by a decrease in correlation length, thus diminishing the group's collective responsiveness, leaving it more vulnerable to predation events. Importantly, in groups undergoing collective turns, the information about the change in direction propagates linearly in time among group members, thanks to the non-reciprocal nature of the visual interactions between individuals. Merging speeds up the transfer of information within each cluster by several fold, while fragmentation slows it down. Our findings suggest that flow interactions may have played an important role in group size regulation, behavioral adaptations, and dispersion in living animal groups.

\spacing{1.5}


\section*{Introduction}

Nature is in a perpetual state of reorganization.
However, while these cohesive patterns are regularly documented in systems of small or moderate size~\cite{Couzin2002,Gautrais2012,Calovi2014,Filella2018,Huang2024,Heydari2024,Peterson2024}, it is unclear how they scale with increasing group size~\cite{Rieucau2015}: do large groups remain cohesive or do they undergo dynamic reorganization? We address this question in massive simulations of schooling fish, where individual swimmers interact through self-generated flows and follow behavioral rules inferred directly from experimental data in shallow water environments~\cite{Gautrais2012,Calovi2018,Filella2018,Huang2024}. By optimizing our computational algorithms, we simulate over long times the motion of groups of up to 50,000 fish. We show that ``more is different"~\cite{Anderson1972,Strogatz2022}. Where smaller groups maintain cohesive and polarized formations, larger groups spontaneously reorganize, constantly fragmenting, scattering and reassembling, similarly to empirical observations of large flocks of birds~\cite{Cavagna2010,Bialek2012} and schools of fish~\cite{Partridge1981,Niwa1998,Rieucau2015,Rieucau2014}. We analyze how this self-reorganization influences the collective responsiveness and speed of information propagation between members of the group~\cite{Couzin2002, Cavagna2010, Attanasi2014}. 

Collective responsiveness in self-organized animal groups manifests in long-ranged spatial correlations~\cite{Nagy2010,Cavagna2010,Ioannou2011}. Correlation measures how the change in the behavior of one individual influences the behavior of others in the group. 
An animal group exhibits maximal responsiveness to a perturbation, say, caused by an attacking predator~\cite{Handegard2012,Rieucau2015}, when 
correlations are scale-free, that is, when the range of spatial correlations scale with the linear group size~\cite{Cavagna2010,Handegard2012}. Analysis of empirical data of large bird flocks confirms that spatial correlations scale linearly with group size $L$~\cite{Cavagna2010}. But do these results translate to groups of swimmers?

In physical models of flow-coupled swimmers, microscopic~\cite{Tsang2014,Tsang2016,Tsang2018} and inertial~\cite{Tsang2013,Zhu2014,Heydari2024,Newbolt2024}, perturbations get amplified as they propagate via the fluid medium, hindering group cohesion. These models do not enable individual swimmers to sense flows and respond accordingly. Biological swimmers, on the other hand, are flow sensitive~\cite{Engelmann2000,Ristroph2015,Colvert2016} 
and seem to correlate their tailbeat frequencies and phase~\cite{Ashraf2017,Zhang2019,Li2020}, but the extent of flow-mediated correlations is limited in space~\cite{Heydari2024}. Recent evidence suggests that vision is both necessary and sufficient for polarized schooling~\cite{Mckee2020}.
Even in robotic agents, visual interactions with immediate neighbors are sufficient to induce scale-free correlations in polarized groups~\cite{Zheng2024}. But are these scale-free correlations universal to groups of individuals with long-ranged visual and hydrodynamic interactions? 
If so, how does dynamic reorganization within the group, including splitting and merging event, affect the extent of spatial correlations? Importantly, how fast does information travel within a polarized group?

Inspired by the analysis of information propagation in bird flocks~\cite{Attanasi2014}, we consider the behavior of our interacting swimmers during spontaneous collective turns. We find that the information about the change in direction propagates linearly in time across the group, at speeds much faster than the individual swimming speed. This is in sharp contrast to the diffusive information propagation in symmetric, consensus-based models~\cite{Vicsek1995}, and in the absence of behavioral inertia~\cite{Attanasi2014}.  We show that symmetry is broken due to the non-reciprocal nature of the interactions between individual swimmers~\cite{Avni2023,Fruchart2021}, much like in the game of telephone, where a player secretly shares a phrase with the next person, who then passes it along to the next player and so on. In this game, the interaction range is one, and the correlation length  — representing how far the phrase spreads before becoming distorted — goes well beyond one, but does not scale with group size~\cite{Mesoudi2008,Carlson2018}. Importantly, the message is transmitted from one person to the next person who did not already have the information. This non-reciprocity inherently breaks symmetry and ensures that the message travels ballistically in time in one direction, 
as opposed to the diffusive propagation that occurs when each person randomly chooses to transmit the information in either direction~\cite{Vicsek1995,Attanasi2014}, left or right, irrespective of where it came from~\cite{Berg1993}. 
Surprisingly, when extending this analysis to quantify the speed of information propagation during self-reorganization, we find that merging of separate clusters speeds up the transfer of information within each cluster by several folds, while splitting and fragmentation slows it down.


\section*{Results}

\begin{figure*}[!t]
\centering
\includegraphics[scale=1]{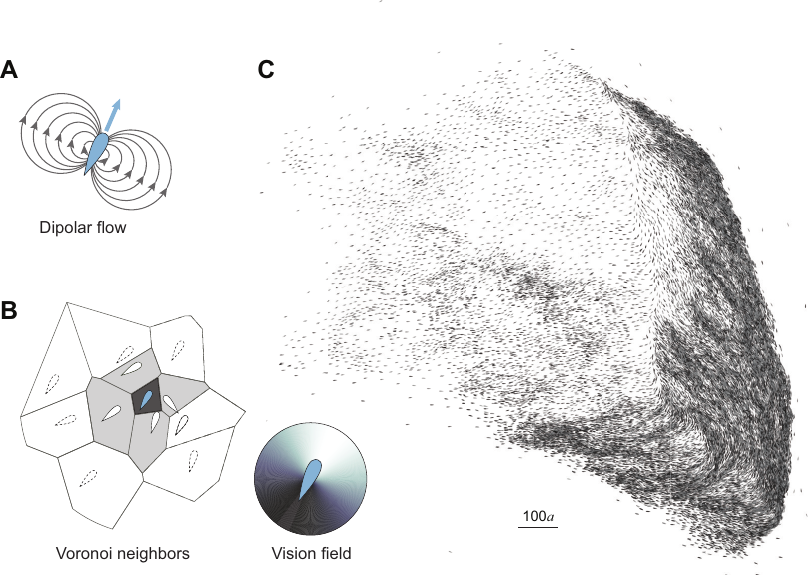}
\caption[]{\footnotesize \textbf{Emergent behavior in a school of 50,000 fish.}  
An individual swimmer \textbf{A.}  creates a dipolar flow disturbance, with dipole intensity proportional to its speed and cross-sectional area, and
\textbf{B.} responds by turning towards and aligning with its first-level Voronoi neighbors, highlighted in grey in this sample Voronoi tesselation. The individual response is mediated by an asymmetric visual field with frontal bias. 
\textbf{C.} School organizes into coherent polarized clusters that dynamically split and merge, exhibiting large density fluctuations, as shown here in a massive merging event involving about 20\% of the fish. In all simulations, total integration time: $T=1000$. Parameter values: $I_a=9$, $I_n=0.5$, $I_f=0.01$, and $N=$ 50,000. Suppl. Movie 1. 
}
\label{fig:50000}
\end{figure*}

\begin{figure*}[!th]
\centering
\includegraphics[scale=1]{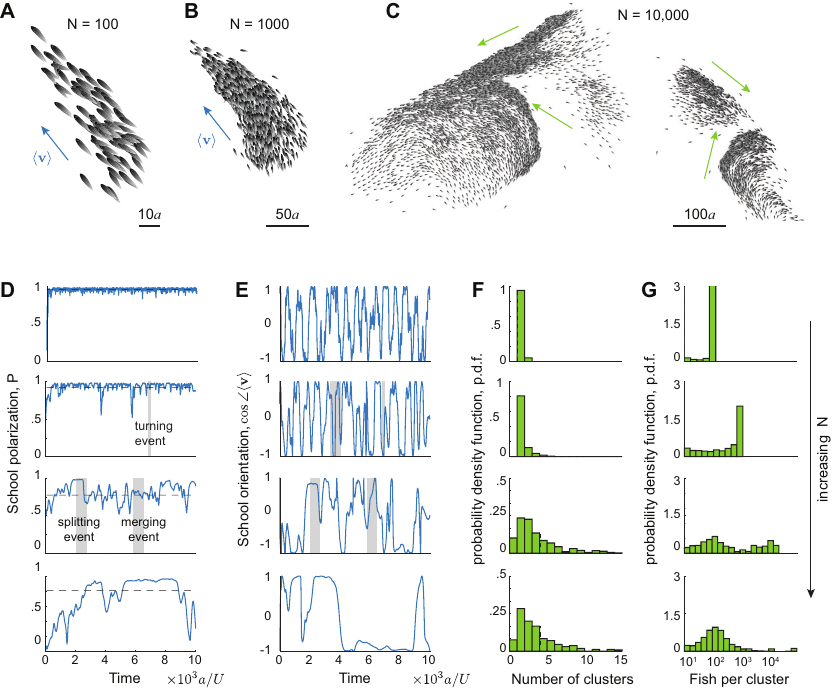}
\caption[]{\footnotesize \textbf{More is different: self-organized behavior depends on group size.} Snapshots of three schools of \textbf{A.} 100, \textbf{B.} 1000, and \textbf{C.} 10,000 fish. For $N$=100 and 1000, the school is globally polarized and remains coherent in time, while for $N=$ 10,000, the school continuously reorganizes, dynamically splitting and merging. Blue arrows indicate the school's average velocity, and green arrows indicate the average velocity of each cluster. Time evolution of \textbf{D.} school polarization $P$ and \textbf{E.} average orientation $\cos \angle \langle \mathbf{v}\rangle$. Distributions of \textbf{F.} number of clusters and \textbf{G.} number of fish per cluster shown in log scale. Parameter values: $I_a = 9$, $I_n = 0.5$, $I_f = 0.01$. In D-G, from top to bottom, $N$ = 100 1000, 10,000, and 50,000. See Suppl. Movies 1 \& 2.  
}
\label{fig:stats}
\end{figure*}

\paragraph{Dynamic reorganization, fragmentation, dispersal and reassembly in large fish schools.}
We numerically simulated the motion of a school of 50,000 fish coupled via visual feedback rules and flow interactions in an unbounded planar domain (Fig.~\ref{fig:50000}, Suppl. Movie 1). Each swimmer followed behavioral rules, modulated by an asymmetrical visual field representing frontal-biased perception~\cite{Gautrais2012}.
These rules were derived empirically from shallow-water experiments, where each swimmer turned towards its Voronoi neighbors, aligned its heading with the same neighbors, and experienced rotational white noise~\cite{Gautrais2012,Calovi2014}. Additionally, each swimmer generated a dipolar flow field and responded to the combined flow generated by all other swimmers~\cite{Filella2018,Huang2024}.
We normalized the swimming speed $U$ and intensity of rotational attraction by a proper choice of characteristic time and length scales~\cite{Filella2018}. Accordingly, with $U=1$, three dimensionless parameters $(I_n,I_a,I_f)$ distinguished the behavior of individual swimmers representing, respectively, the rotational noise, alignment, and hydrodynamic intensities (Methods). By definition, the hydrodynamic intensity $I_f$ introduces an additional dimensionless length scale $a= \sqrt{I_f/U}$ that reflects the swimmer's bodylength. Here, we used parameter values $(I_n,I_a,I_f)$ that, in smaller groups of $100$ fish, led to stable polarized schooling~\cite{Filella2018,Huang2024} (Fig.~\ref{fig:stats}A). We optimized our computational algorithms in order to scale our simulations to groups of the order of $10^4$ swimmers (Methods). In the group of 50,000 fish, starting from random initial conditions, the fish self-organized into coherent polarized structures that dynamically fragmented and reassembled, exhibiting large density fluctuations (Fig.~\ref{fig:50000}, Suppl. Movie 1), comparable to empirical observations of large bird flocks~\cite{Cavagna2010} and fish schools~\cite{Partridge1981,Niwa1998, Rieucau2015,Rieucau2014}.

\paragraph{More is different.} 
We systematically varied the number of swimmers $N$. In Fig.~\ref{fig:stats}A-C, we report  cohesive and highly polarized schools of 100 and 1000 swimmers and loss of global cohesion in a school of 10,000 swimmers, where distinct polarized clusters moved in different directions.
Statistical results from sample simulations at $N=$ 100, 1000, 10,000, and 50,000 are reported in Fig.~\ref{fig:stats}D-G. The polarization order parameter $P = |\sum_{j=1}^N e^{i \theta_j}|/N$, where $\theta_j$ is the orientation of swimmer $j$, is consistently close to 1 for $N=$ 100 and 1000, indicating high polarization at all time.
For $N$= 10,000 and 50,000, $P$ fluctuates violently, reflecting the reorganization and constant splitting and merging in larger schools: a sharp decrease in $P$ indicates a splitting event, while a sharp increase indicates a merging event. 

Considering the velocity $\langle\mathbf{v}\rangle = \sum_{j=1}^N \mathbf{v}_j/N$ of the entire school, we found that, on average, schools swam faster than the individual swimming speed $U$ for $N=$100 and 1000, consistent with~\cite{Filella2018}, but slower for $N=$10,000 and 50,000 because of the break-up of these larger schools into subgroups that themselves swam faster but in random directions. For example, in the snapshots in Fig.~\ref{fig:stats}A-C, the school moved at an average speed of $1.20$, $1.08$, and $0.54$ for $N=$100, 1000 and 10,000, respectively, that is, at nearly two-fold slower than the individual speed $U=1$ for $N=$10,000. The highly-polarized clusters that formed within the larger schools could reach equally high speeds as their free counterparts; for example, in Fig.~\ref{fig:stats}C, while the overall speed of the school was $0.54$, the four clusters moved at speeds 1.14, 
0.83, 0.86, 1.08, albeit in different directions  (Suppl. Movie 2).
The time evolution of $\cos(\angle\langle \mathbf{v} \rangle)$, where $\angle\langle \mathbf{v} \rangle$ represents the school's overall orientation, shows more frequent changes in orientation at smaller $N$, whereas in the larger schools, frequent splitting and merging events create subgroups that move in random directions, hindering the entire school from turning together cohesively. 
Fig.~\ref{fig:stats}F and G show the number of subgroups per school identified by a density-based clustering algorithm (Methods, \cite{Ester1996,Campello2013,Mcinnes2017}) and the average number of fish per cluster. The larger schools at $N$= 10,000 and 50,000 exhibited wider distributions, reminiscent of empirical observations~\cite{Niwa1998},
suggesting the existence of a capacity of number of swimmers per polarized cluster that follows a distribution skewed towards moderate values, with a heavy tail beyond which the cluster breaks up and reorganizes. Because of the behavioral and statistical similarities between $N$= 10,000 and $N=$ 50,000, and to save computational effort, hereafter we investigate the mechanisms responsible for this behavior in groups of up to 10,000 fish.

\begin{figure*}[!t]
\centering
\includegraphics[scale=1]{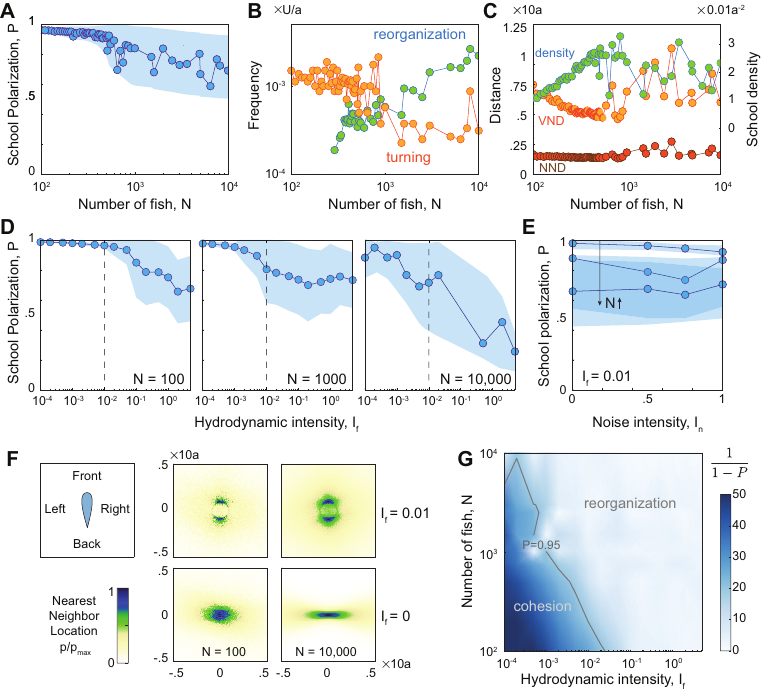}
\caption[]{\footnotesize \textbf{School cohesiveness depends on the hydrodynamic intensity of individual swimmers.} 
 \textbf{A.} time-averaged polarization $P$ versus school size $N$ indicates a transition from a highly-polarized cohesive regime to a regime of constant dynamic organization beyond a critical group size; shaded area indicates standard deviation of $P$ within time series; $P$ is averaged over the last 80\% of the simulation time, discounting the initial 20\% to eliminate transient effects. \textbf{B.} Dominant frequency of ${d P}/{dt}$ and $\cos \angle \langle \mathbf{v} \rangle $ versus school size show an increasing frequency of splitting and merging, reflected by sharper changes in $dP/dt$, with increasing $N$,  \textbf{C.} average nearest neighbor distance (NND), average distance to Voronoi neighbors (VND), and average density. In A-C, hydrodynamic intensity is set to $I_f = 0.01$;
corresponding plots without hydrodynamic interactions ($I_f=0$) are shown in Fig.~\ref{figsi:nohydro}. 
 Time-averaged polar order parameter $P$ and standard deviation as a function of \textbf{D.} hydrodynamic intensity $I_f$ 
 and \textbf{E.}  noise intensity $I_n$ for schools of size  $N=$100, $N=$1000,  and $N=$10,000.  
\textbf{F.} Heatmap of nearest neighbors for $N=100$, and $N=10,000$. Top row: with hydrodynamic interaction $I_f=0.01$; bottom row: without hydrodynamic interactions $I_f=0$. 
\textbf{G.} Instead of $P$, we plot ${1}/({1-P})$ to enhance the contrast of the colormap over the space of hydrodynamic intensity $I_f$ and number of swimmer $N$. Results show loss of cohesion with increasing $N$ and $I_f$. Parameter values: $I_a = 9$, $I_n = 0.5$. 
}
\label{fig:size_hydro}
\end{figure*}

In Fig.~\ref{fig:size_hydro}A, we report the time-averaged values of the school polarization $P$ as a function of $N$. As we varied $N$ from 100 to 10,000,
up to $N\approx$ 1000, the swimmers exhibited stable schooling, behaving mostly as an indivisible entity, with consistently high polarization values $P$ greater than $0.95$.
Beyond $N = 1000$, the school began to fragment, forming locally polarized subgroups that dynamically rejoined and separated again. This indicates the existence of a bifurcation depending on school size, past which the dynamic reorganization within the school caused a decrease in the global polarization order parameter and an increase in its variance (Fig.~\ref{fig:size_hydro}A). 
In the highly polarized and cohesive regime, the school turned frequently and rarely fragmented, but as $N$ increased, the frequency of global turning events decreased while the frequency of splitting and merging increased (Fig.~\ref{fig:size_hydro}B).
The average density of the school increased monotonically up to $N\approx 1000$, while, locally, the average nearest neighbor distance (NND) remained nearly unchanged and the average distance to Voronoi neighbors (VND) decreased (Fig.~\ref{fig:size_hydro}C). That is, in the cohesive regime, the school became denser with increasing $N$, not by getting uniformly closer to all neighbors, but by getting closer to distant neighbors while maintaining the same distance to nearest neighbors, consistently with experimental observations~\cite{Peterson2024}. As $N$ increased beyond the cohesive regime, the average density and distance to nearest and Voronoi neighbors (NND and VND) all exhibited large fluctuations, reflecting a transition to a new regime of dynamic reorganization within the school. 

\paragraph{Flow interactions trigger spontaneous reorganization within the school.} We next asked what mechanisms lead to school self-reorganization at larger $N$. Given that our model accounts for vision-based rules of alignment and attraction, flow interactions, and individual noise, we set out to test  
 the role of each in triggering the transition from the cohesive state to the state of self-reorganization with increasing school size. We first suppressed all hydrodynamic interactions, and considered a school of 10,000 swimmers interacting only via vision-based rules. We observed no fragmentation, reassembly, and reorganization, independent of noise levels (Supplementary Fig.~\ref{figsi:nohydro}D). At exceedingly large noise, the school transitioned to a swarming phase where all polarization was lost, consistent with classic models~\cite{Couzin2002,Couzin2003,Couzin2005, Calovi2014, Filella2018}. 
We thus concluded that the vision-based rules of attraction and aligning to Voronoi neighbors lead to no fragmentation of the group, independent of group size, and that noise alone is not sufficient for self-reorganization. 
Without hydrodynamic interactions, the average density of the school increased monotonically with the number of swimmers, leading to unrealistically dense patterns and distributions of nearest neighbor distance (Fig.~\ref{fig:size_hydro}E) that do not fit with experimental observations~\cite{Peterson2024}.
Hydrodynamic interactions are important. We next maintained the same noise level and varied the intensity of the hydrodynamic interactions by increasing the dipolar field $I_f$ across several orders of magnitudes from $10^{-4}$ to 5: since $I_f \sim a^2U$ is proportional to the swimmer's speed $U$ and the square of the bodylength $a$, a weaker dipolar intensity represents smaller and slower fish and a larger dipolar intensity represents larger and faster fish~\cite{Filella2018}. In Fig.~\ref{fig:size_hydro}D, we report results across this wide range of $I_f$ for $N=$ 100,  1000, and 10,000 swimmers. Smaller schools maintain school cohesion at larger values of $I_f$. In larger schools, cohesion is lost at smaller values of $I_f$, indicating that the capacity for cohesive schools depends on the hydrodynamic intensity of individual swimmers, which in turn depends on their size and speed. That is, smaller fish can school cohesively in larger numbers. To confirm our findings that flow interactions drive self-reorganization, we found that with hydrodynamic interaction and without noise, the phenomena of dynamic reorganization, fragmentation, dispersal, and reassembly remain largely unchanged (Fig.~\ref{figsi:nonoise}). 
Thus, in the context of our model, 
hydrodynamic interactions are both necessary and sufficient for self-reorganization.

\paragraph{Scale-free correlation breaks down during school self-reorganization.} 
The range of spatial correlations in polarized flocks of birds was shown to scale with the maximal length of the flock~\cite{Cavagna2010}. 
This linear scaling of correlation length with group size implies that the effective perception range of each individual encompasses the entire group and enables transfer of information between members regardless of distance, ensuring collective response to perturbations~\cite{Cavagna2010,Bialek2012,Rieucau2014}. 
We asked whether these conclusions are generic to emergent polarization in groups of self-propelled individuals, including our simulations of schooling fish, and how self-reorganization within the school, in the form of continuous fragmentation, dispersal, and reassembly, affects the range of spatial correlation and the ability to transfer information among school members.

\begin{figure*}[!t]
\centering
\includegraphics[scale=1]{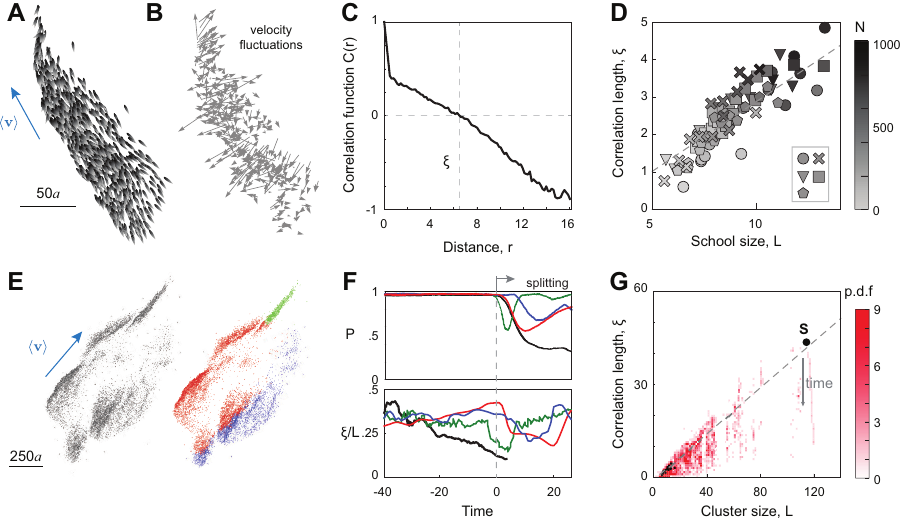}
\caption[]{\footnotesize \textbf{Scale-free correlations in velocity fluctuations are compromised during school reorganization and fragmentation.}  \textbf{A.} A snapshot of a stable and cohesive school of $N = 1000$ swimmers and \textbf{B.} corresponding velocity fluctuations. \textbf{C.} Correlation function $C(r)$ (\eqref{eq:correlation_velofluc}) in velocity fluctuations between pairs swimmers as a function of their mutual distance $r$. 
\textbf{D.} The correlation length $\xi$ is linear in school size $L=\text{max}\|\mathbf{x}_i -\mathbf{x}_j\|$, with $\xi \approx 0.37 L - 0.84$, and coefficient of determination $R^2 = 0.83$, for all alignment and noise intensities in the cohesive and polarized regime; 
here, $(I_a,I_n)$ are given by $\circ: (9,0.5)$, $\times: (7,0.5)$, $\nabla: (5,0.5)$, $\square: (9,0.3)$, \pentagon: $(9,0.7)$.
Greyscale indicates the number of fish $N$. 
\textbf{E.} Snapshots of school with $10,000$ fishes prior to split. The right panel shows the coloring based on the clusters after splitting. 
\textbf{F.} Polar order parameter $P$ and correlation length over school size $\xi/L$ for the whole group (black) and individual clusters (red, blue, and green with corresponding colors). 
\textbf{G.} Histogram of correlation length and cluster size for all clusters emerging in the simulation of $10, 000$ fish. 
}
\label{fig:correlation_school}
\end{figure*}

To address these questions, we considered cohesive and highly polarized groups of swimmers ranging in size from $N$ = 100 to 1000, where consistent with~\cite{Cavagna2010}, we analyzed snapshots with high degree of polarization ($P>0.9$). 
For swimmer $i$, we defined the fluctuation $\delta \mathbf{v}_i$ around the group's mean velocity as $\delta \mathbf{v}_i= \mathbf{v}_i- \langle\mathbf{v}\rangle$  (Fig.~\ref{fig:correlation_school}A,B). By construction, $\sum_{i=1}^N\delta \mathbf{v}_i= 0$, indicating no net fluctuations in the net motion of the center of mass of the school. We calculated the spatial correlation function $C(r)$ of velocity fluctuations (Methods), 
where the span of $r$ does not exceed the length $L$ of the group defined as $L=\max\|\mathbf{x}_i-\mathbf{x}_j\|$.
A positive value of $C(r)$ close to 1 implies that the fluctuations are nearly parallel and strongly correlated. Conversely, a negative value of $C(r)$ close to $-1$ implies that the fluctuations are antiparallel and anticorrelated. A value of $C(r)\approx 0$ implies a random distribution of velocity fluctuations with no correlation. In Fig.~\ref{fig:correlation_school}C, we report $C(r)$ versus $r$ for the snapshot presented in~Fig.~\ref{fig:correlation_school}A.  At short distances, the correlation is close to 1 and decays with increasing $r$, becoming negative at large interindividual distances, indicating strong correlation at short distances and strong anticorrelation at large distances, and in no range of $r$ are the velocity fluctuations uncorrelated.

To explain the behavioral implications of this form of $C(r)$, we defined the correlation length $\xi$ as the relative distance $r$ at which $C(\xi)=0$. By definition, the value of $\xi$ is the maximal size of the positively correlated domain.  
In Fig.~\ref{fig:correlation_school}D, the resulting correlation length $\xi$ is plotted versus school length $L$ using simulations at various sets of parameters $(I_f, I_a,I_n)$ and school size $N\leq 1000$, provided $P>0.9$ (Table~\ref{tab:summary}). We found that $\xi$ increases linearly with $L$, much like in the case of starling flocks~\cite{Cavagna2010}. 
We found no scale-free behavior in speed correlations, because speed fluctuations in our model are due to passive hydrodynamic interactions and do not arise from active interactions between the swimmers (Fig.~\ref{figsi:correlation_speed}).
These results confirm that scale-free correlations in velocity fluctuations are generic. They reflect the rotational interactions encoded at the level of individual swimmers (Methods, Eqs.~\eqref{eq:eom}), and can be attributed to the existence of a Goldstone mode associated with the breaking of rotational symmetry leading to group polarization~\cite{Cavagna2018,Bialek2022}. 
Interestingly, in our simulations, the slope of the best fit of $\xi$ vs $L$ is nearly one-third, similar to the slope reported in~\cite{Cavagna2010} for natural bird flocks. 

But does this scale-free correlation generalize to larger groups that continuously reorganize? To answer this question, 
we revisited the simulation of $N=$ 10,000 fish reported in Fig.~\ref{fig:stats} and identified cohesive and highly polarized clusters within the school that are about to undergo self-reorganization. For example, in Fig.~\ref{fig:correlation_school}E, we report a snapshot where the entire school moves cohesively, at high polarization, preceding a splitting event, where the school fragments into three different clusters (highlighted in different colors).
We calculated the time evolution of the polarization parameter $P$ of the entire school and of the subgroups that later constituted the three separate clusters (Fig.~\ref{fig:correlation_school}F). The school maintained a high level of polarization until the time at which it fragmented, beyond which $P$ decreased, but each cluster recovered quickly exhibiting high polarization per cluster. 
Interestingly, a gradual decrease in the correlation length $\xi$ far preceded the sharp decrease in $P$, while the school size $L$ remained unchanged (Fig.~\ref{figsi:turn_curvature}E), inducing an overall decrease in $\xi/L$ over time and loss of scale-free correlation prior to fragmentation. This loss in scale-free correlation is predictive of an upcoming splitting event in all cohesive clusters. 

To verify this, we considered the time evolution of the school of 10,000 swimmers and, at each snapshot, we identified all clusters of cohesive swimmers, selected highly-polarized clusters for which $P>0.9$, calculated the corresponding $\xi$ and $L$, and plotted the joint probability density function of cluster size $L$ and correlation length $\xi$ as a heatmap over the $(L,\xi)$ space (Fig.~\ref{fig:correlation_school}G). The $(L,\xi)$ values are concentrated at and below the scale-free correlation line (dashed grey) obtained in stable schools in Fig.~\ref{fig:correlation_school}D.
Highlighted on this plot are the $(L,\xi)$ values corresponding to the fragmentation event reported in  Fig.~\ref{fig:correlation_school}E,F: the correlation length starts at the scale-free line $\xi/L \sim 1/3$ and decreases before the onset of splitting (grey arrow), emphasizing the loss of scale-free correlation during school reorganization.

\begin{figure*}[!t]
\centering
\includegraphics[scale =1]{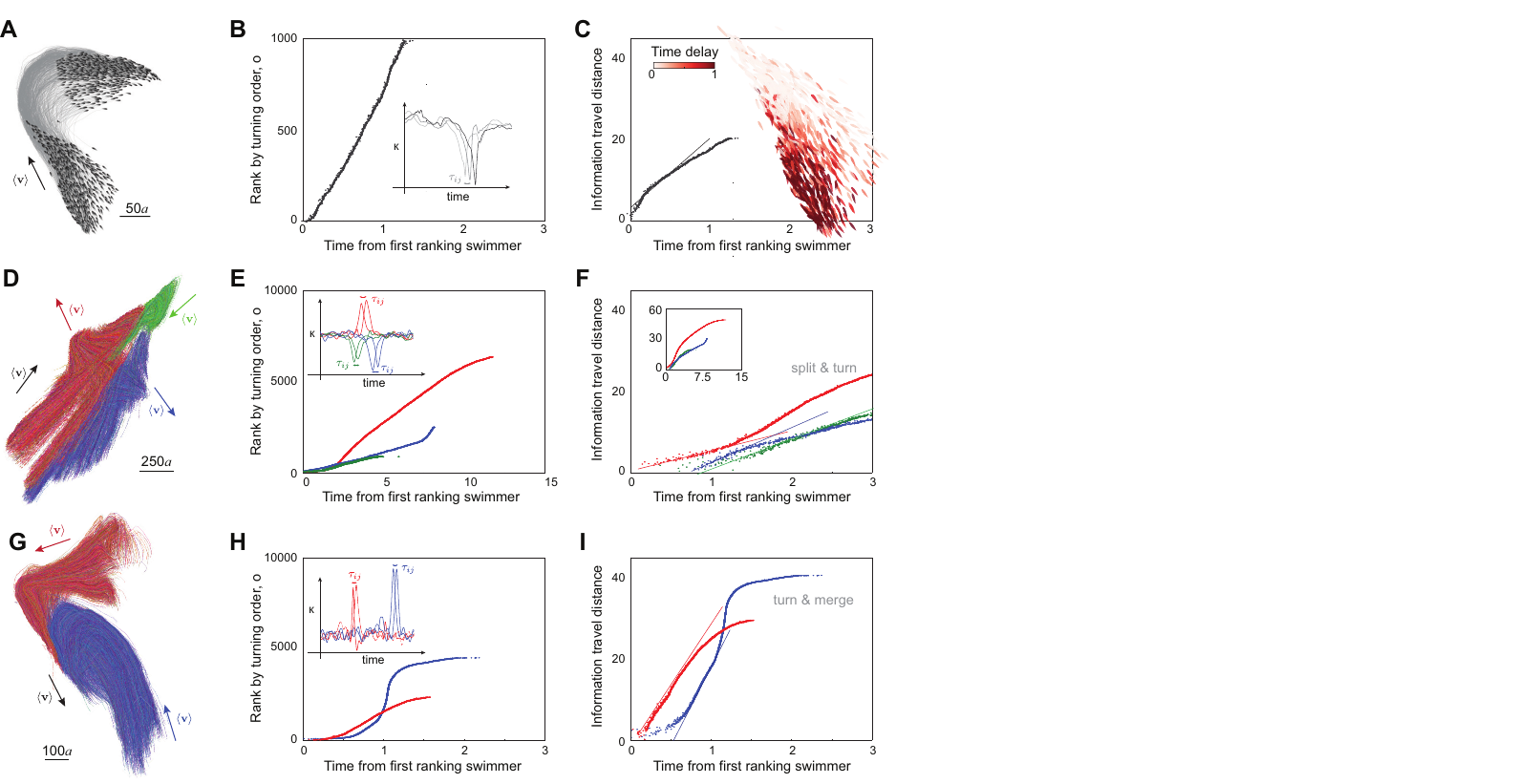}
\caption[]{\footnotesize \textbf{Information transfer during turning, splitting and merging.} \textbf{A.} Polarized school of $N = 1000$ swimmers turns spontaneously by ``free will". 
\textbf{B.} Rank of fish by the order $o$ at which they reach maximal curvature (inset) and \textbf{D.} information travel distance defined as $\sqrt{o/\text{density}}$ versus absolute turning time delay; slope indicates that information travels linearly in time at speed equal to $17$ times the individual swimming speed $U$. Inset shows absolute turning time as a colormap over all swimmers.
 $P$ and  $\xi /L$ are reported in Fig.~\ref{figsi:turn_curvature}B. 
\textbf{D.} Trajectories of individual fish in a polarized school of $N = $ 10,000 swimmers that later split into three clusters highlighted in red, blue, and green (Fig.~\ref{fig:correlation_school}E,F). 
\textbf{E.} Rank of fish within each cluster by the order $o$ at which they reach maximal curvature (inset) and \textbf{F.} information travel distance versus the absolute turning time delay; information travels at slower speeds compared to freely turning at speeds of 5, 9, and 7 times $U$ for the red, blue, and green subgroups, respectively. 
Colormap of time delays, $P$ and $\xi/L$ are reported in Fig.~\ref{figsi:turn_curvature}.
\textbf{G.} Trajectories of individual fish showing the merging of two clusters highlighted in red and blue in a school of  $N = $ 10,000 fish.
\textbf{H.} Rank of fish within each cluster by the order $o$ at which they reach maximal curvature (inset) and  \textbf{I.} information travel distance versus absolute turning time delay; information travels at slower speeds compared to freely turning at speeds of 30 and 39 times $U$ for the red and blue subgroups, respectively. 
Colormap of time delays, $P$ and $\xi/L$ are reported in Fig.~\ref{figsi:turn_curvature}G.
For a slow-motion replay of these events, see Suppl. Movie 3.  
}
\label{fig:turn}
\end{figure*}

\paragraph{Information propagates linearly thanks to the non-reciprocal visual interactions between swimmers.}
Scale-free correlation reflects the potential for indirect transfer of information between individuals in the group but it does not describe the efficiency of a collective response to environmental factors~\cite{Cavagna2010,Attanasi2014}. An efficient collective response depends on how localized perturbations succeed in modifying the behavior of the entire group. Take, for example, a group changing its overall heading direction (Fig.~\ref{fig:turn}A and Suppl. Movie 3). The actual execution of such turns cannot be instantaneous, because a certain amount of time is needed to propagate the turn throughout the group. During this time, cohesion is strained by the mismatch between individuals who have already turned and those who have not yet done so, as reflected by a drop in polarization $P$ (Fig.~\ref{figsi:turn_curvature}B). 
Therefore, the speed at which information is transferred from individual to individual plays a crucial role in maintaining group cohesion, which in return is key for scale-free correlation and collective responsiveness.

We set out to quantify information transfer in cohesive groups first, then to assess the effect of school reorganization -- fragmentation and merging -- on information transfer. To fix ideas, we analyzed, following~\cite{Attanasi2014}, a collective turn in a cohesive group of $N = 1000$ swimmers. Given the full trajectory of each swimmer $i$ in the group (Fig.~\ref{fig:turn}A), we calculated the curvature $\kappa_i$ as a function of time and identified the time $t_i$ of maximum curvature. For each pair of swimmers, $i$ and $j$, we calculated their mutual turning delay, $\tau_{ij}=t_i-t_j$, defined as the amount of time by which swimmer $j$ turns before $(\tau_{ij}>0)$ or after $(\tau_{ij}<0)$ swimmer $i$ (Methods, Fig.~\ref{fig:turn}B, inset, and \ref{figsi:turn_curvature}C). From the delays $\tau_{ij}$, we ranked all swimmers in the group according to their turning order, identifying the first to turn, the second, and so on. We then labeled each swimmer $i$ by its order $o_i$ in terms of its absolute turning time $t_i$ with respect to the top-ranking swimmer. We found that the top-ranking swimmers -- the first swimmers to turn -- are physically close to each other (Fig.~\ref{fig:turn}C, inset). That is, the collective turn has a spatially localized origin that propagates across the group through swimmer-to-swimmer transfer of information.   

\begin{figure*}[!t]
\centering
\includegraphics[scale = 1]{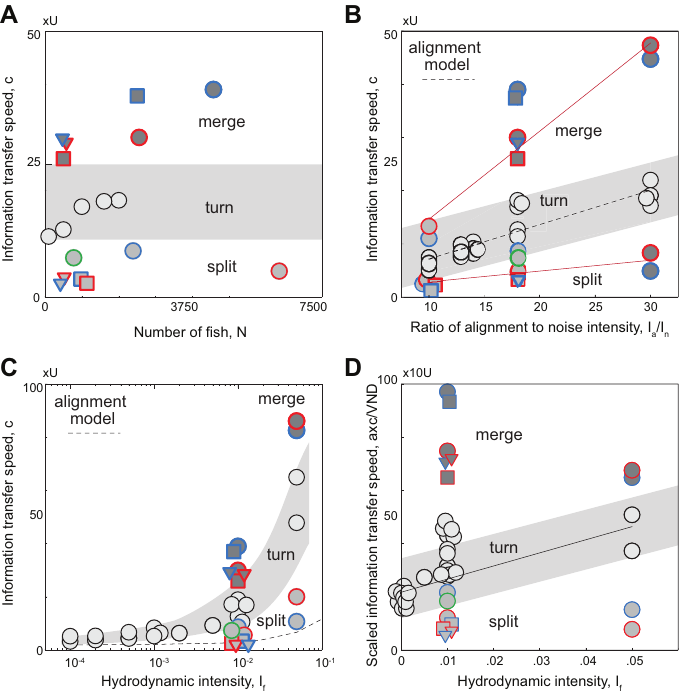}
\caption[]{\footnotesize \textbf{Flow interactions speed up information transfer.} Information transfer speed $c$ (in units of  $U$)  \textbf{A.}  shows weak dependence on number of swimmers $N$ during turning, splitting, merging. 
\textbf{B.} $c$ scales linearly with $Ia/I_n$, with $c=0.65I_a/I_n+0.73$ at $R^2=0.92$ during free turning,  $c=0.20I_a/I_n+0.84$ at $R^2=0.98$ during splitting, and $c=1.65I_a/I_n-1.91$ at $R^2=0.96$ during merging. 
Results obtained for the same parameter sets $(I_a,I_n)$ used in Fig.~\ref{fig:correlation_school}. Fig.~\ref{figsi:IaIn}A-C shows that polar order parameter $1/(1-P) $ scales linearly with $I_a/I_n$~\cite{Attanasi2014}.  
\textbf{C.} $c$
increases with increasing hydrodynamic intensity $I_f$. 
The dashed line shows the prediction of information transfer speed $c$ using the non-reciprocal alignment model $c \sim I_a \alpha $ where $\alpha$ is taken to be the average VND. 
\textbf{D.}  
Because hydrodynamic intensity $I_f$ affects VND (Fig.~\ref{figsi:IaIn}D), we subtract this effect  by scaling $c$ with $VND/a$. If $I_f$ affects $c$ only through its effect on VND, we expect the scaled information speed to be independent of $I_f$. Instead, it  increase linearly with $I_f$ during free turning, with $c/\langle \textrm{VND}\rangle =492.46 I_f +21.79$ at $R^2=0.77$, indicating that flow interactions increase information travel speed. Parameter values: $I_a=9$ and $I_n=0.5$. 
}
\label{fig:turn_compare}
\end{figure*}

Given this ranking, we sought to describe how much distance $d$ the information travels in a time $t$. Given that the motion of the group is two-dimensional and that the turn has a localized origin, the information propagates a distance  $d_i=\sqrt{o_i/\rho}$, where $\rho$ is the school density which remains nearly constant during the turn~\cite{Attanasi2014}. Plotting $d_i$ versus time (Fig.~\ref{fig:turn}C), we found a clear linear regime at early and intermediate times, implying that, following the first-rank fish, the distance traveled by the information grows linearly with time $d(t) = c t$, where $c$ is the speed of propagation of information; its value is about $20$ times that of the self-propelled speed $U$ of individual swimmers in our model. We repeated this analysis for various turning instances in schools ranging in size from $100$ to $2000$ swimmers (Fig.~\ref{figsi:turn}). The information transfer speed fluctuated with the number of swimmers but remained consistently an order of magnitude larger than that of the individual swimming speed (Fig.\ref{fig:turn_compare}A).

The linear and fast propagation of information within the school is a key factor in preserving school cohesion during turning. What is the mechanism responsible for this phenomenon? Theoretical models based on local alignment with neighboring individuals, such as the Vicsek model~\cite{Vicsek1995}, lead to diffusive information propagation, with speeds that scale sublinearly with $\sqrt{t}$~\cite{Attanasi2014}.  
The discrepancy between the diffusive model predictions and the linear information travel speeds obtained in empirical data of flocks of birds was attributed to the presence of inertia and associated conservation law~\cite{Attanasi2014}. However, our results are based on a kinematic model~\eqref{eq:eom} that ignores inertia of individual swimmers; thus, accounting for inertia is not necessary for the information to travel linearly in time.  The crucial factor in our model is the non-reciprocal visual interactions between individuals. Indeed, we derived a continuum partial differential equation governing the phase $\varphi$, where $\varphi_i =\theta_i-\langle \theta\rangle$ is the perturbation from the school average heading direction $\langle \theta \rangle = \angle\mathbf{v}$ (Methods). We found that due to the non-reciprocity induced by visual interactions, information propagates linearly from the front to the back of the school at speed $c \propto  I_a  \alpha$, where $\alpha$ is a characteristic, average distance to neighbors. We disregarded noise in deriving this scaling law, assuming that the alignment intensity is dominant. To test this prediction in simulations with noise, we systematically varied both the alignment and noise intensities and calculated the resulting polarization $P$ in cohesive groups (Fig.~\ref{figsi:IaIn}). We found that $P$ satisfies the relation $P = 1 - I_n/I_a$ derived in~\cite{Attanasi2014} using the spin-wave approximation. We also calculated the information transfer speed $c$ during turning (Fig.~\ref{fig:turn_compare}B) as a function of $I_a/I_n$ and found that, indeed, $c$ scales linearly with $I_a/I_n$, demonstrating consistency between our simulations and the scaling law derived from the alignment model.
Our findings complement those of~\cite{Attanasi2014}, showing that non-reciprocal visual interactions lead to information transfer speeds that scale linearly with time, without the need to invoke inertia. Yet, our results differ from \cite{Attanasi2014} in two ways: $c$ scales linearly with alignment intensity $I_a$, in contrast to the sublinear scaling $c  \propto  \sqrt{I_a}$ in \cite{Attanasi2014}. Importantly, our theory predicts an anisotropy in the information transfer speed, with information traveling faster in the longitudinal direction of the school, while in~\cite{Attanasi2014}, the information transfer speed is isotropic. Assessment of the relative effects of inertia versus non-reciprocal visual interactions on the transfer of information in natural animal groups would require integrating models with empirical data~\cite{Rosenthal2015}. 

\paragraph{Fragmentation slows down information propagation and merging speeds it up.} We next examined splitting events during school self-reorganization. In Fig.~\ref{fig:turn}D, we show trajectories of the splitting event pointed out earlier (Figs.~\ref{fig:stats}D, \ref{fig:correlation_school}E), where the school of $10,000$ swimmers, starting from a polarized state, splits into three subgroups (labeled in red, blue, and green), with each subgroup turning in a different direction. We analyzed each subgroup, computing the turning sequence of each swimmer within their subgroup (Figs.~\ref{fig:turn}D, \ref{figsi:turn_curvature}D-F) and calculated the information travel speed within each subgroup (Fig.~\ref{fig:turn}F).
The different subgroups have nearly the same information transfer speed, about three fold the self-propelled velocity, which is much slower compared to free turning (Fig.~\ref{fig:turn}F, \ref{fig:turn_compare}A). This is perhaps not surprising given the loss of spatial correlation with fragmentation (Fig.~\ref{fig:correlation_school}F,G).
The change in information transfer speed can also be attributed to the presence of an attraction term. While attraction to the neighboring group is beneficial during merging, it is detrimental to information transfer during splitting (Method). 

Lastly, we examined information transfer during merging events. In Fig.~\ref{fig:turn}G, we show trajectories of a merging event in the simulation of 10,000 swimmers, where two subgroups (labeled in red and blue), starting from polarized states in different directions, turn and merge into a single subgroup. 
During merging, swimmers from different subgroups do not mix; the two subgroups turn individually, move closer and reach consensus on a joint moving direction. 
Again, we analyzed each subgroup, computing the turning sequence of each swimmer relative to its own subgroup, and calculated the information travel speed within each subgroup (Figs.~\ref{fig:turn}E,F, \ref{figsi:turn_curvature}G-I).
Interestingly, the information transfer speed increases, reaching up to $40$ fold that of the self-propelled speed, much faster than information propagation in free turning and during fragmentation. To further probe the robustness of these results, we analyzed multiple merging events in clusters of different sizes ranging from 1000 to 3000 (Fig.~\ref{figsi:merge}). We found that the information transfer speed is consistently higher during merging, indicating that continuous information input from neighboring clusters increases the speed of information propagation (Fig.~\ref{fig:turn_compare}).

\paragraph{Flow interactions enhance information travel speeds.}
We next explored the effect of flow interactions on information travel speed. In~Fig.~\ref{fig:turn_compare}C, we fixed the alignment and noise intensities and systematically varied the hydrodynamic intensity $I_f$ from $10^{-4}$ to $0.05$. Results are shown in Fig.~\ref{fig:turn_compare}C on a semi-log scale. At small hydrodynamic intensity, the school stays cohesive and the information transfer speed follows closely that predicted by the vision-based alignment model $c \propto  I_a  \alpha$, where $\alpha$ is taken to be equal to the average VND. However, as $I_f$ increases, the information speed diverges from this model prediction. 

To help explain the effect of hydrodynamic interactions on information propagation, we considered how a perturbation in phase $\varphi_i$ propagates via the fluid medium only (Methods). In the continuum limit, we found that, as in the alignment model, hydrodynamic interactions alone cause information to travel from the front to the back of the school with speed $c \propto I_f/\alpha^2$. However, this flow-based scaling does not correctly predict the information travel speed of the school, because of the non-trivial interplay between vision and hydrodynamic interactions.
Indeed, hydrodynamics affects the average distance to Voronoi neighbors (Fig.~\ref{fig:stats}F), which in turn affects the information travel speed due to alignment. If hydrodynamic coupling between swimmers had no direct effect on information propagation other than through its effect on the average VND, we would expect the information travel speed normalized by the average VND to be independent of $I_f$; rather, it increases linearly with $I_f$ (Fig.~\ref{fig:turn_compare}D), indicating that hydrodynamic interactions, coupled to vision-based alignment, enhance information transfer speeds.


\section*{Discussion}

We explored information propagation in mathematical models of massive schools of fish, consisting of up to 50,000 individuals. We showed that (1) as the school size increases, flow interactions destabilize global polarization, creating locally polarized clusters that dynamically self-organize, fragment and reassemble, akin to empirical observations in natural fish schools~\cite{Niwa1998}; (2) while correlations in velocity fluctuations in cohesive and polarized clusters are scale-free, splitting events are preceded by a decrease in correlation length; (3) information propagates linearly in time within cohesive groups, at speeds exceeding 20 times that of the swimming speed of the individual, thanks to the non-reciprocal nature of visual interactions between individuals, with inertia playing no part in this ballistic information transfer speed; (4) the speed of information propagation is robust to group size but varies with self-organization: merging of separate clusters increases the speed of information transfer within each cluster, while fragmentation decreases it; and (5) flow interactions enhance the information propagation speed.

Our findings have important implications on size regulation~\cite{Higashi1993,Bonabeau1999} and behavioral adaptations~\cite{Roberts1996} in living animal groups. 
Our model predicts that larger and faster-moving swimmers that generate stronger dipolar flows fragment with increasing number of swimmers, and smaller-sized swimmers can school cohesively in bigger numbers; Indeed, several of the natural species of fish that form massive schools, such as sardines, herring, and anchovies, have relatively small bodylength, not exceeding 30 cm, and typical swimming speeds of 1 to 2 bodylength per second. The prospect that flow physics may have played a role in the evolution and regulation of group size is an exciting direction for future work~\cite{Rieucau2015}. 

Another key area to explore in future work is the role of flow interactions in modulating the spatial dispersal of fish species~\cite{Bowler2005}. In our model, flow interactions cause large schools to disperse in random directions, akin to a `divide and conquer' strategy, where the group splits up and explores different regions of the space independently before regrouping. Our results are consistent with observations in pelagic fish schools that fragment and rejoin, with many species of fish exhibiting spatial distributions that are skewed toward small sizes with a long tail toward large sizes~\cite{Niwa1998}. Here, we went beyond reporting the fragmentation-rejoining process to proposing a flow-based mechanism that drives, or at least contributes, to this important behavior in natural fish schools. Understanding the factors that influence spatial dispersal patterns is important because these patterns, in turn, influence numerous  processes that are fundamental for the survival of population, such as mate-finding~\cite{Carde1997,Huang2017,Hang2023,Jiao2024}, disease transmission~\cite{Kent2009,Mittal2020}, foraging and prey-detection~\cite{Pohlmann2001,Berdahl2013,Basil1994,Weissburg1994,Handegard2012}, and predator avoidance~\cite{Radakov1973, Gerlotto2006,Jiao2023}. 

In natural animal groups preyed upon by faster-moving predators~\cite{Rieucau2014}, the speed of information propagation within the group is critical to ensure a swift response to predatory threat. Our result that school fragmentation -- a strategy thought to confuse predators~\cite{Humphries1970,Couzin2003} -- comes at the cost of decreasing information propagation speed within the group, suggests an evolutionary trade-off between maximizing information propagation within the group and creating confusion for the predator. It also suggests that fragmenting the school could be an effective predation strategy that weakens the perception range of the prey, especially in collective predation~\cite{Angelani2012, Papadopoulou2022}.

\paragraph{Acknowledgement.} Funding support provided by the NSF grants RAISE IOS-2034043 and CBET-210020, ONR grants N00014-22-1-2655 and N00014-19-1-2035, and NIH grant R01-HL153622 (all to E.K.).

\paragraph{Authors contribution.} 
E.K. conceptualized and supervised the research; H.H. and C.H. wrote code with input from A.B.; H.H. performed simulations and collected data; H.H. and E.K. analyzed the data and prepared figures; E.K. wrote the manuscript and all authors edited and approved it.

\clearpage
\vfill
\newpage

\printbibliography

\section*{Methods}

{\footnotesize
\subsection*{A. Mathematical model of individual swimmers}
We consider a system of $N$ fish, where each fish is represented as a self-propelled particle moving at a constant speed $U$ (m$\cdot$s$^{-1}$) relative to the flow velocity. A fish creates a flow disturbance represented by its far-field potential dipole~\cite{Tsang2013,Kanso2014} and follows behavioral laws derived from shallow water experiments~\cite{Gautrais2012,Calovi2014,Filella2018,Huang2024}. Accordingly, each fish interacts with the local flow generated by all other fish and reorients its heading direction to both get closer and align with its Voronoi neighbors~\cite{Filella2018,Huang2024}. Consider that fish $i$ is located at $\mathbf{x}_i \equiv (x_i, y_i)$ in an inertial $(x,y)$-frame, with velocity $\mathbf{v}_i = \dot{\mathbf{x}}_i$, where $\dot{()}$ represents derivative with respect to time $t$, and has a heading direction $\mathbf{p}_i \equiv (\cos \theta_i, \sin \theta_i)$ expressed in terms of a heading angle $\theta_i$ measured from the $x$-axis. We write the equations of motion of fish $i$ directly in non-dimensional form, using the length scale $\sqrt{U/k_p}$ and timescale $1/\sqrt{Uk_p}$, where $k_p$ (m$^{-1}\cdot$s$^{-1}$) is the intensity with which a fish reorients to get closer to its neighbors~\cite{Filella2018},
\begin{equation} 
\begin{aligned}
\dot{\mathbf{x}}_i= U \mathbf{p}_i+ \mathbf{U}_i, \qquad
d \theta_i= \langle r_{ij} \sin\theta_{ij} + I_{a} \sin{\phi_{ij}}\rangle dt + \Omega_i dt +I_n d W_t.
\end{aligned}
\label{eq:eom}
\end{equation}
Here, speed is normalized to $U=1$. The non-dimensional noise intensity $I_n$ scales a standard Wiener process $W(t)$ modeling the fish ``free will"~\cite{Wiener1976}. 
The term $\langle\circ\rangle$ represents the fish reorientation in response to visual feedback: it means that fish $i$ only ``sees" its Voronoi neighbors $\mathcal{V}_i$, with attraction intensity normalized to one and non-dimensional alignment intensity $I_a$, both averaged with weight $1+\cos \theta_{ij}$ modeling continuously a rear blind angle~\cite{Calovi2014},
\begin{equation} 
\begin{aligned}
\langle\circ\rangle=\sum_{j \in \mathcal{V}_i} \circ\left(1+\cos \theta_{i j}\right) / \sum_{j \in \mathcal{V}_i}\left(1+\cos \theta_{i j}\right).
\end{aligned}
\label{eq:vision}
\end{equation}
The intermediate variables $r_{ij} = \| \mathbf{x}_i - \mathbf{x}_j\|$, $\theta_{ij} = \left(\angle (\mathbf{x}_j - \mathbf{x}_i) -\theta_i\right)$, and $\phi_{ij} = \theta_j-\theta_i$ represent, respectively, the relative distance, viewing angle, and difference in heading angle between fish $i$ and $j$. The vector  $\mathbf{{U}}_i$ represents the flow velocity generated by all other swimmers at the location of swimmer $i$ and $\Omega_i$ denotes the angular velocity 
\begin{equation} 
\begin{aligned}
\mathbf{U}_i =\sum_{j=1, j\neq i}^N \frac{I_f}{\pi} \frac{\mathbf{p}_j^{\perp} \sin 2\theta_{ji}+\mathbf{p}_j \cos 2\theta_{ji}}{r^2_{ij}} 
, \qquad \Omega_i = \mathbf{p}_i \cdot \frac{d\mathbf{U}_i}{d\mathbf{x}}|_{\mathbf{x}_i} \cdot \mathbf{p}_i^{\perp},
\end{aligned}
\label{eq:dipole}
\end{equation}
where $I_f=\pi (a/2)^2 U$ is the strength of fish-induced dipolar flow field, with $a$ indicating the fish bodylength and $\mathbf{p}^\perp$ is a unit vector orthogonal to $\mathbf{p}$~\cite{Kanso2014}. $I_f=0.01$ gives $a=0.11$ in dimensionless form. 
Eqs.~\eqref{eq:eom}--~\eqref{eq:dipole} form a closed set of $3N$ differential equations governing the $3N$ unknowns $(x_i,y_i,\theta_i)$, where $i=1,\ldots,N$. These equations depend solely on three non-dimensional parameters, $I_n$, $I_a$, and $I_f$ representing the noise, alignment, and hydrodynamic intensities. 

\subsection*{B. Computational method} 
To numerically solve the system of equations~\eqref{eq:eom} for a large number of fish $N$, one needs a computationally efficient approach to handle the all-to-all hydrodynamic interactions and Voronoi tessellation at each time step. 
The computational complexity due to the hydrodynamic interactions in Eq.~\ref{eq:dipole} scales with $\mathcal{O}(N^2)$.
To handle these interactions, we optimized and paralleled the code responsible for computing the direct sum in Eq.~\ref{eq:dipole}  using a just-in-time compiler called~\textit{Numba}~\cite{Lam2015}. \textit{Numba} compiles, optimizes, and parallelizes the Python code to approach the computational performance of C or Fortran. 
Note that fast multipole methods (FMM) reduce the computational complexity of the hydrodynamic interactions from $\mathcal{O}(N^2)$ to nearly $\mathcal{O}(N)$~\cite{Greengard1987,Ying2004}, but
FMM algorithms do not have a significant advantage over direct sum in systems of the order of $10^4$ agents~\cite{Ying2004}, hence our choice to directly optimize the $\mathcal{O}(N^2)$ sum in~\eqref{eq:dipole}.
For the Voronoi tessellation in two dimensions (2D), efficient algorithms exist for reducing this task to $\mathcal{O}(N \log N)$~\cite{Barber1996}. We utilized the function~\texttt{Delaunay} in \textit{Scipy}~\cite{2020SciPy}.
We implemented these approaches in evaluating the right-hand sides of Eq.~\eqref{eq:eom} at each time step $dt$, discretized the noise term using $dW_t = \mathcal{N}(0,1)\sqrt{dt}$, and used an explicit Euler–Maruyama method to integrate~\eqref{eq:eom} forward in time at a timestep size $dt= 10^{-2}$~\cite{Kloeden1992}. We run our algorithm on an Exxact Valence Workstation with a 56-core Intel Xeon W9-3495X CPU. With this software and hardware setup, a timestep takes about $1$ second for $10,000$ agents, with hydrodynamic interactions and Voronoi tessellation taking about half of the computational time each. 
Integrating the motion of $10,000$ agents over a time interval $T=1000$ took about a day; integrating the motion of $50,000$ swimmers for the same time interval took about three weeks.

\subsection*{C. Statistical and data-driven analysis} 

\paragraph{Polar order parameter.} To quantify the degree of polarization within each group or subgroup of swimmers, we calculated the polar order parameter $P = \|\sum_{j=1}^N \mathbf{p}_j\|/N \in [0,1]$, where 
$P = 1$ when all swimmers are heading in the same direction; it is nearly zero for randomly oriented swimmers. 

\paragraph{Identifying splitting and merging events} Fish remained cohesive in relatively small groups, but in large schools, we observed dynamic splitting and merging where the large school got divided into subgroups, each moving in a different direction that seemed to randomly rejoin and divide again for the entire simulation time. To identify these splitting and merging events, we examined the time evolution of the polar order parameter: $P$ rapidly decreased or increased when a splitting or emerging event occurred. To determine the time scale at which these events took place, we calculated the dominant frequency of  $dP/dt$ using Fast Fourier transformation (FFT). In the absence of splitting and merging events, such as at small number of fish, the FFT is characterized by high frequencies due to individual-level noise. We discarded these frequencies (equivalent to a low-pass filter) to identify the frequencies at which the splitting and merging events occurred in large schools. We discard all frequencies larger than $0.5$. The inverse of this dominant frequency defines the time scale of splitting and merging. 

\paragraph{Clustering algorithm.}  To identify the number of distinct subgroups in large groups of swimmers as a function of time, we used a numerical approach based on clustering methods~\cite{Miyahara2023}. Because in this active system, the individual clusters have versatile and time-varying shapes, we needed a computational approach that could handle arbitrarily shaped clusters. Classic clustering methods based on expectation–maximization algorithms~\cite{Dempster1977}, such as K-means~\cite{Macqueen1967} or Mixture Models~\cite{Duda1973}, suffer in identifying intertwined clusters with time-varying shapes. Here, we used density-based methods that are designed to separate low- and high-density regimes in the domain and identify complex-shaped clusters~\cite{Ester1996,Campello2013,Mcinnes2017,Schubert2017,Kang2018};  particularly, we used the Hierarchical Density-Based Spatial Clustering of Applications with Noise (HDBSCAN) algorithm~\cite{Ester1996,Campello2013,Mcinnes2017}, implemented in the \textit{scikit-learn} package~\cite{Pedregosa2011}, which has been successfully applied to identify clusters in simulations of the Vicsek model~\cite{Miyahara2023}. 

\paragraph{Spatial correlation in velocity fluctuations.} The degree of polarization $P$ provides little insights into the collective response in a school~\cite{Cavagna2010,Cavagna2018}. To understand the collective response, we examined how fluctuations in each swimmer's velocity correlate with those of others.
 For swimmer 
$i$, we defined the fluctuation $\delta \mathbf{v}_i$ around the group's mean velocity as $\delta \mathbf{v}_i= \mathbf{v}_i- \langle\mathbf{v}\rangle_{N}$, where $\langle\mathbf{v}\rangle_N = \sum_{j=1}^N \mathbf{v}_j/N$. By construction, $\sum_{i=1}^N\delta \mathbf{v}_i= 0$, which simply indicates no net motion in the center of mass reference frame of the school. We defined the spatial correlation function $C(\mathbf{r})$ of fluctuations, which measures the average inner product of velocity fluctuations of swimmers at a distance $r$ from each other,
\begin{equation} 
\begin{aligned}
C(r) = \dfrac{1}{C_{o}} \dfrac{\sum_{i}\sum_j (\delta \mathbf{v}_i \cdot \delta \mathbf{v}_j) \delta (r-r_{ij} )}{ \sum_{i}\sum_j \delta (r-r_{ij})}. 
\end{aligned}
\label{eq:correlation_velofluc}
\end{equation}
Here, the Dirac-delta function $\delta(r-r_{ij})$, where $r_{ij}=\|\mathbf{r}_{ij}\|$ and $\mathbf{r}_{ij} = \mathbf{x}_i-\mathbf{x}_j$, selects pairs of swimmers at mutual distance $r$, 
and $C_{o}$ is a normalization factor such that $C(r=0)=1$. 

\paragraph{Time delays during turning and information propagation within the group.} When a cohesive polarized group of swimmers performed a collective turn, to define the turn, we examined the time evolution of the curvature $\kappa_i = \dfrac{\| \mathbf{v}_i\times \dot{\mathbf{v}}_i\|}{\| \mathbf{v}_i \|^3}$ of the trajectory traced by swimmer $i$, where $\dot{\mathbf{v}}_i$ is the swimmer's acceleration. In 2D, the curvature can be calculated directly in terms of the time derivatives of the coordinates $(x_i,y_i)$,  namely, $k_i(t)=\dfrac{\dot{x}_i\ddot{y}_i-\dot{y}_i\ddot{x}_i}{(\dot{x}_i^2+\dot{y}_i^2)^{3/2}}$.
The time-evolution of the curvature $\kappa_i(t)$  of a swimmer $i$ undergoing a turn exhibits a maximum at the time of the turn. Inspired by~\cite{Attanasi2014,Attanasi2015}, and given two swimmers $i$ and $j$, we defined the mutual turning delay $\tau_{ij} $ as the time required to shift the full curve of $\kappa_j(t)$ to maximally overlap it with $\kappa_i(t)$
\begin{equation} 
\begin{aligned}
\tau_{ij}=\argmax_{\tau} k_i(t)k_j(t-\tau).
\end{aligned}
\label{eq:time_delay}
\end{equation}
Here, $\tau_{ij} < 0$ means fish $i$ turns ahead of fish $j$ and vice versa. In the absence of noise, time ordering requires that $\tau_{ij} = \tau_{ik} + \tau_{kj}$, for each triplet $i$, $j$, $k$. For example, if $i$ turns 10 time units before $k$, and $k$ turns 5 time units before $j$, then $i$ turns 15 time units before $j$. Because we are dealing with a noisy system, this equality may not be strictly satisfied, but $\tau_{ij}$ is equal to $\tau_{ik} + \tau_{kj}$ on average.

We next ranked the group of fish undergoing a turn based on their time of maximal curvature. For each fish $i$, we calculated how many other fish it has turned ahead of~\cite{Attanasi2014, Cavagna2014}. The order of this number -- the number of other fish a fish precedes in turning -- defines a rank for the fish; the first-ranked fish is ahead of the largest number of fish and its turning time is used to set the time $t_1$ of the onset of the turning event. In a perfect system, with no noise, the turning time $t_i$ of a lagging fish $i$ can be calculated directly relative to the turning time of the first-ranked fish $1$, $t_i = t_1 + \tau_{i1}$. However, because the system is noisy, this method of calculating $t_i$ introduces small statistical errors. To minimize these errors, we define $t_i$ using the mutual delay $\tau_{ij}$ with respect to all swimmers $j$ ranked higher than $i$,
\begin{equation} 
t_i=\frac{1}{\text{{rank}}_i-1}\sum_{\text{\tiny{rank}}_j < \text{\tiny{rank}}_i}(t_i+\tau_{ij}), \qquad i > 1
\label{eq:absolute_turning_time}
\end{equation}

\subsection*{D. Coarse-grained analysis of information propagation}

\paragraph{Alignment model.} Starting from the microscopic equation describing the time evolution of swimmer's heading
\begin{equation} 
\begin{aligned}
\dot\theta_i=  I_{a} \frac{\sum_{j\in\mathcal{N}_i}\sin{\phi_{ij}}(1+\gamma\cos\theta_{ij})}{\sum_{j\in\mathcal{N}_i}(1+\gamma\cos\theta_{ij})},
\end{aligned}
\label{eq:eom_derive}
\end{equation}
where $\gamma \in[0,1]$ is a parameter that controls the strength of vision-based bias, or non-reciprocity, toward neighbors in front: $\gamma=1$ is used in~\eqref{eq:eom} while $\gamma=0$ means no visual bias. We derive a continuum equation under the following conditions. Firstly, we consider a highly polarized school, which means that the orientation of each swimmer within the school can be decomposed into the average heading direction of the school $ \langle \theta \rangle$ and a small fluctuation $\varphi_i$ of individual swimmers $i$ about the average heading $\theta$, namely $\theta_i= \langle \theta \rangle + \varphi_i$. Without loss of generality, we assume the $ \langle \theta \rangle =0$, which aligns the positive $x$-direction with the moving direction of the group.
Based on this assumption, $\sin \phi_{ij} =\sin(\theta_j-\theta_i) =\sin(\varphi_j-\varphi_i) \approx \varphi_j-\varphi_i$, and $\cos\theta_{ij} = \cos (\arctan\frac{y_j-y_i}{x_j-x_i}-\theta_i) = \cos (\arctan\frac{y_j-y_i}{x_j-x_i}-\varphi_i)$. Substitute these relationships into~\eqref{eq:eom}, we get
\begin{equation} 
\begin{aligned}
\frac{\partial \varphi_i}{\partial t}=  \frac{I_{a}}{N} \sum_{j\in \mathcal{N}_i}(\varphi_j-\varphi_i)\left[1+\gamma\cos (\arctan\frac{y_j-y_i}{x_j-x_i}-\varphi_i)\right],
\end{aligned}
\label{eq:eom2}
\end{equation}
Secondly, we assume that the swimmers are located on a 2d lattice of mesh size $\alpha$, and mesh orientation aligned with the swimming direction. We aim to coarse-grain the discrete equations~\eqref{eq:eom} over a coarse-graining box containing a focal swimmer and four immediate neighbors, such that a swimmer $i$ responds to its direct front, left, back, and right neighbors, indexed by $i1$, $i2$, $i3$, $i4$. Their locations with respect to particle $i$ can be written as 
$\mathbf{x}_{i1}-\mathbf{x}_{i}= (\alpha,0)$, $\mathbf{x}_{i2}-\mathbf{x}_{i}= (0,\alpha)$, $\mathbf{x}_{i3}-\mathbf{x}_{i}= (-\alpha,0)$, and  $\mathbf{x}_{i4}-\mathbf{x}_{i}= (0,-\alpha)$.
Plug it into \eqref{eq:eom2} and reorganize to arrive at
\begin{equation} 
\begin{aligned}
\frac {\partial \varphi_{i}}{\partial t} = \frac{ \alpha^2I_a }{4}  
\left( \frac{\varphi_{i1} +\varphi_{i3} -2\varphi_{i}}{\alpha^2} + \frac{\varphi_{i2} +\varphi_{i4} -2\varphi_{i}}{\alpha^2} \right) 
+ \frac{ \gamma \alpha I_a }{2}  
\left(   \cos \varphi_{i} \frac{\varphi_{i1}  -\varphi_{i3}}{2\alpha} + \sin \varphi_{i} \frac{\varphi_{i2}  -\varphi_{i4}}{2\alpha}  \right)
\end{aligned}
\label{eq:2dunbiased_derive2}
\end{equation}
The finite difference can be approximated by first-order and second-order derivatives, such that at small $\varphi$ where $\cos \varphi \sim 1$, $\sin \varphi \sim \varphi$, we arrive at
\begin{equation} 
\begin{aligned}
\frac {\partial \varphi}{\partial t} &=   \frac{\alpha^2 I_a }{4}  \Delta \varphi+ \frac{\gamma \alpha I_a}{2} \left( \frac{\partial \varphi}{\partial x} +\varphi \frac{\partial \varphi}{\partial y}  \right)
\end{aligned}
\label{eq:2dbiased_derive3}
\end{equation}
The equation governing $\phi$ has an anisotropic advection term, where $\phi$ is advected linearly in the longitudinal direction $x$ and non-linearly, albeit at much smaller speed (considering that $\varphi \ll 1$) in the transverse direction. Ignoring the nonlinear term, we get
\begin{equation} 
\begin{aligned}
\frac {\partial \varphi}{\partial t} &=   \frac{\alpha^2 I_a }{4}  \Delta \varphi+ \frac{\gamma \alpha I_a}{2}  \frac{\partial \varphi}{\partial x}.
\end{aligned}
\label{eq:2dbiased_derive4}
\end{equation}
The diffusion term scales with $\alpha^2 I_a$, while the advection term scales with $\alpha I_a$, implying that in dense schools, linear advection is dominant. Ignoring diffusion and considering an initial perturbation in the longitudinal $x$-direction of form $\varphi_0 (x,y)= A\sin(k_x x)$, the perturbation propagates from front to back at a speed  $c= \gamma \alpha  I_a/2$,
\begin{equation} 
\begin{aligned}
\varphi(t,x,y)= A\sin(k_x (x+ c t)).
\end{aligned}
\label{eq:2dbiased_pde_linear_inviscid_x}
\end{equation}

\paragraph{Hydrodynamic interaction model.}  

Considering the group is heading in the same direction and ignoring noise and all vision-based interactions in~\eqref{eq:eom}, a small perturbation in $\varphi_i$ about the heading direction propagates via hydrodynamic interactions only following the simpler equation 
\begin{equation} 
\begin{aligned}
\frac{\partial \varphi_i}{\partial t}=  \mathbf{p}_i \cdot \frac{d\mathbf{U}_i}{d\mathbf{x}}|_{\mathbf{x}_i} \cdot \mathbf{p}_i^{\perp}.
\end{aligned}
\label{eq:hydro_lattice1}
\end{equation}
Here, to simply the analysis, we consider the swimmers to form an infinite one-dimensional lattice with equally-spaced potential dipoles of mesh size $\alpha$, such that the flow field at location $i$ is given by~\cite{Tsang2018},
\begin{equation} 
\begin{aligned}
\mathbf{U}_i =\sum_{j=-\infty, j\neq i}^\infty \frac{I_f}{\pi} \frac{\mathbf{p}_j^{\perp} \sin 2\theta_{ji}+\mathbf{p}_j \cos 2\theta_{ji}}{r^2_{ij}} 
\end{aligned}
\label{eq:hydro_lattice}
\end{equation}
Considering perturbations of wavenumber $k$ and associated wavelength $2\pi/k=(K-1)\alpha$, where $K$ is the perturbed number of swimmers,  we employ the analytical expression derived in~\cite{Tsang2018}, which transforms the infinite summation in~\eqref{eq:hydro_lattice} to a finite summation. Substituting back into~\eqref{eq:hydro_lattice1}, we get
\begin{equation} 
\begin{aligned}
\frac{\partial \varphi_i}{\partial t}=\frac{-2\pi^2I_f}{K^3\alpha^3} \sum_{j=1, j \neq i}^K \frac{\cos \left[ {\pi}(i-j)/{K}\right]}{\sin^3 \left[ {\pi} (i-j)/{K}\right]} \sin (\varphi_j-2\varphi_i)
\end{aligned}
\label{eq:hydro_lattice_derive1}
\end{equation}
Linearizing using $\sin (\varphi_j-2\varphi_i) \approx \varphi_j-2\varphi_i$,
and approximating the finite difference by first-order derivatives, we arrive at
\begin{equation} 
\begin{aligned}
\frac{\partial \varphi}{\partial t}=\frac{2 I_f k}{\pi \alpha}  \frac{\partial \varphi}{\partial x}
\int_{\alpha}^{\pi/2} \frac{x\cos x}{\sin^3 x} dx,
\end{aligned}
\label{eq:hydro_lattice_derive3}
\end{equation}
where the integral is a constant depends only on wave number $k$. This shows that perturbations propagate linearly while getting amplified by hydrodynamic interactions.

\vfill


\clearpage
\newpage

\newpage
\vfill

\setcounter{figure}{0}
\setcounter{table}{0}
\renewcommand{\thetable}{S\arabic{table}}  
\renewcommand{\thefigure}{S\arabic{figure}}

\begin{figure*}[!th]
\centering
\includegraphics[scale=1]{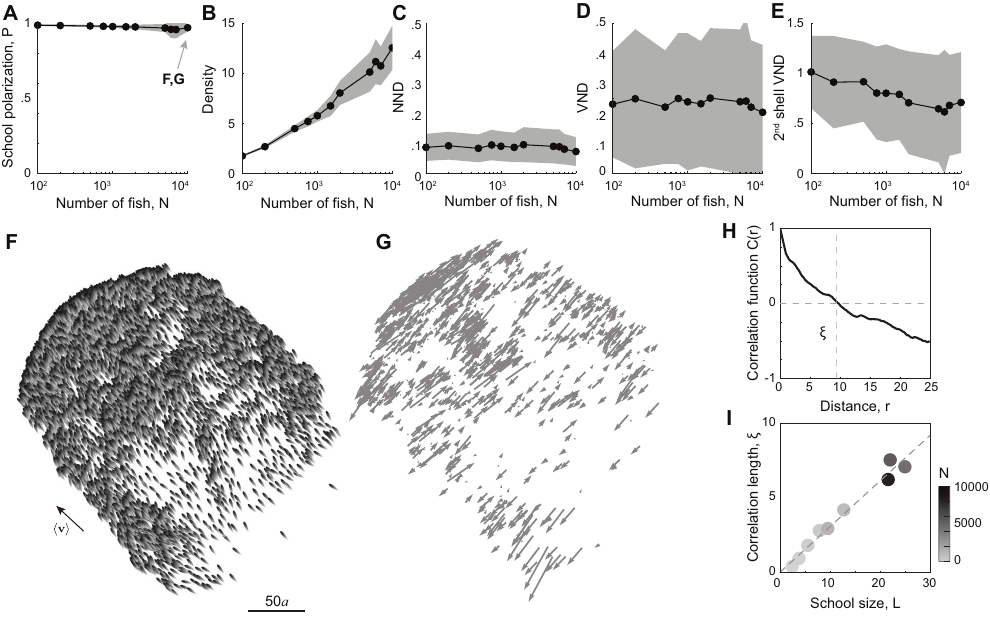}
\caption[]{\footnotesize \textbf{Polarized schools do not split in the absence of hydrodynamic interactions.} 
\textbf{A.} Average polar order parameter $P$ is nearly unchanged with increasing number of swimmers $N$ when $I_f = 0$. 
\textbf{B.} density increases monotonically with increasing number of swimmers.
\textbf{C.} average nearest neighbor and \textbf{D.} average distance to Voronoi neighbors are nearly unchanged with increasing number of swimmers, albeit with larger fluctuations in the latter. 
\textbf{E.} average distance to second shell Voronoi neighbors decreases with increasing number of swimmers.
Snapshots of \textbf{F.} the school composed of $10,000$ swimmers  and \textbf{G.} corresponding velocity fluctuations in the absence of hydrodynamic interactions.  
\textbf{H.} Correlation function plotted as a function of distance for the snapshots in \textbf{F.} and\textbf{G.}. 
\textbf{I.}  Correlation length $\xi$ is a linear function of school size $L$. The slope of the fitting line is $ 0.30$. The slope is close to the slope we got with hydro (Fig.~\ref{fig:correlation_school}D) and in~\cite{Cavagna2010}. In all simulations,  total integration time is $T=1000$.
}
\label{figsi:nohydro}
\end{figure*}

\begin{figure*}[!th]
\centering
\includegraphics[scale=1]{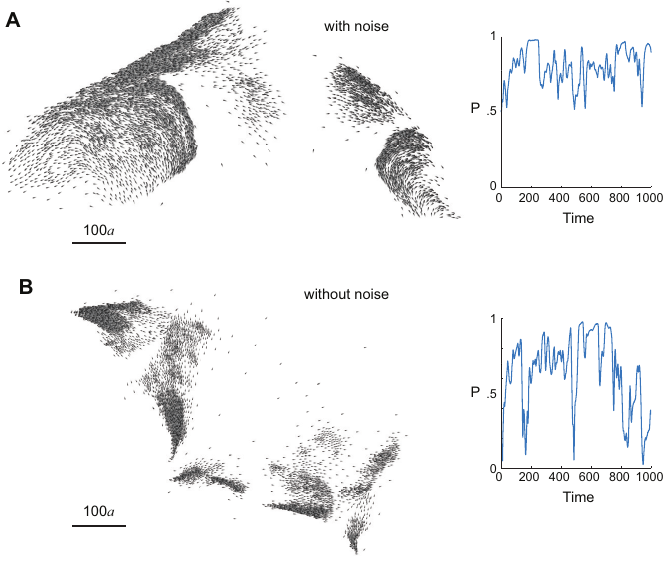}
\caption[]{\footnotesize \textbf{Noise is not necessary for self-organization.} 
\textbf{A.} A snapshot and time evolution of polar order parameter $P$ for a case with $N=10,000$ swimmer with noise ($I_n=0.5$). \textbf{B.} A snapshot and time evolution of polar order parameter $P$  for a case with $N=10,000$ swimmer without noise ($I_n=0$). Parameter values: $N=10,000$, $I_a=9$, and $I_f=0.01$. 
}
\label{figsi:nonoise}
\end{figure*}

\begin{figure*}[!t]
\centering
\includegraphics[scale =1]{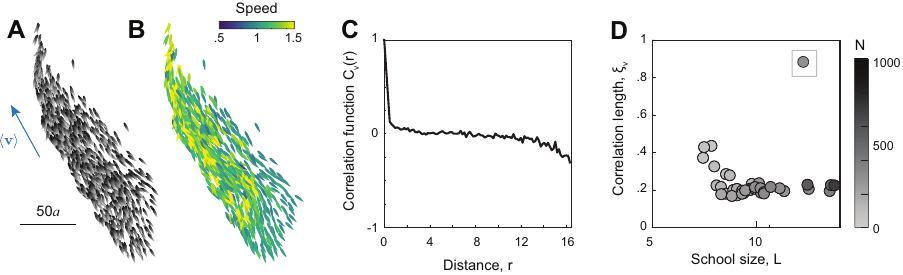}
\caption[]{\footnotesize \textbf{Spatial correlations in speed and corresponding correlation length. } 
\textbf{A.} A snapshot of a stable school with $N = 1000$ and \textbf{B.} speed of individual swimmers plotted as a colormap. \textbf{C.} Correlation function $C_v(r)= 
\left[\sum_{i}\sum_j ( (||\mathbf{v}_i||- \langle||\mathbf{v}||\rangle) (||\mathbf{v}_j||- \langle||\mathbf{v}||\rangle)  \delta (r-r_{ij})\right]/C_{o}\sum_{i}\sum_j \delta (r-r_{ij})$, where $C_o$ is a normalization constant 
is the average product of the speed fluctuations of pairs of fishes at mutual distance $r$ . Fitting $C_v(r)$ to an exponential decay $C_v(r)=A\exp(-r/\xi_v)$ gives a fitted correlation length $\xi_v$. \textbf{D.} Correlation length $\xi_v$ plotted versus school size $L$. 
}
\label{figsi:correlation_speed}
\end{figure*}

\begin{figure*}[!t]
\centering
\includegraphics[scale =1]{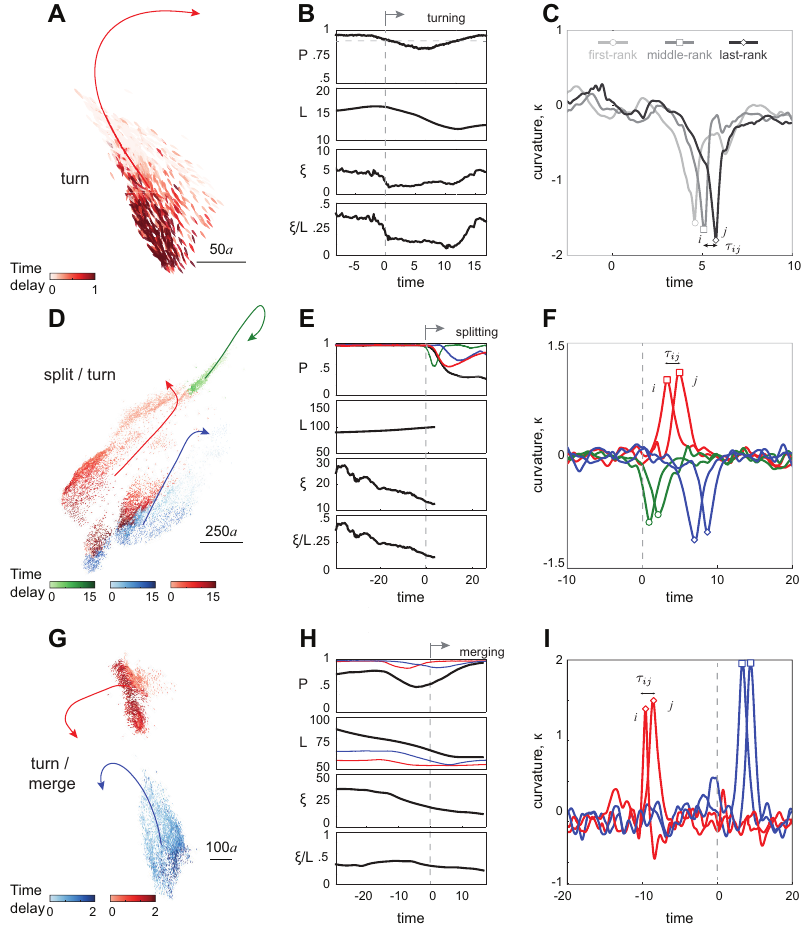}
\caption[]{\footnotesize \textbf{Analysis of turning, splitting, and merging events.} 
\textbf{A.} Absolute turning time plotted as a colormap over the fish school at the onset of turning. \textbf{B.} Polar order parameter $P$, school size $L$, and correlation length $\xi$, $\xi/L$ versus time. \textbf{C.} Sample curvature versus time for first-rank, middle-rank, and last-rank swimmers.
(A-C correspond to the turning event in Fig.~\ref{fig:turn}A). 
\textbf{D.} Absolute turning time plotted as a colormap over the fish school prior to splitting.
\textbf{E.} Polar order parameter, school size, and correlation length for each subgroup versus time. \textbf{F.} Sample curvature from each subgroup versus time. 
(D-F correspond to the splitting event in Fig.~\ref{fig:turn}D). 
\textbf{G.} Absolute turning time plotted as a colormap over the fish school prior to merging. \textbf{H.} Polar order parameter,  school size and correlation length for each subgroup. \textbf{I.} Sample curvature from each subgroup versus time.
(G-I correspond to the merging event in Fig.~\ref{fig:turn}G). 
}
\label{figsi:turn_curvature}
\end{figure*}



\begin{figure*}[!t]
\centering
\includegraphics[scale =1]{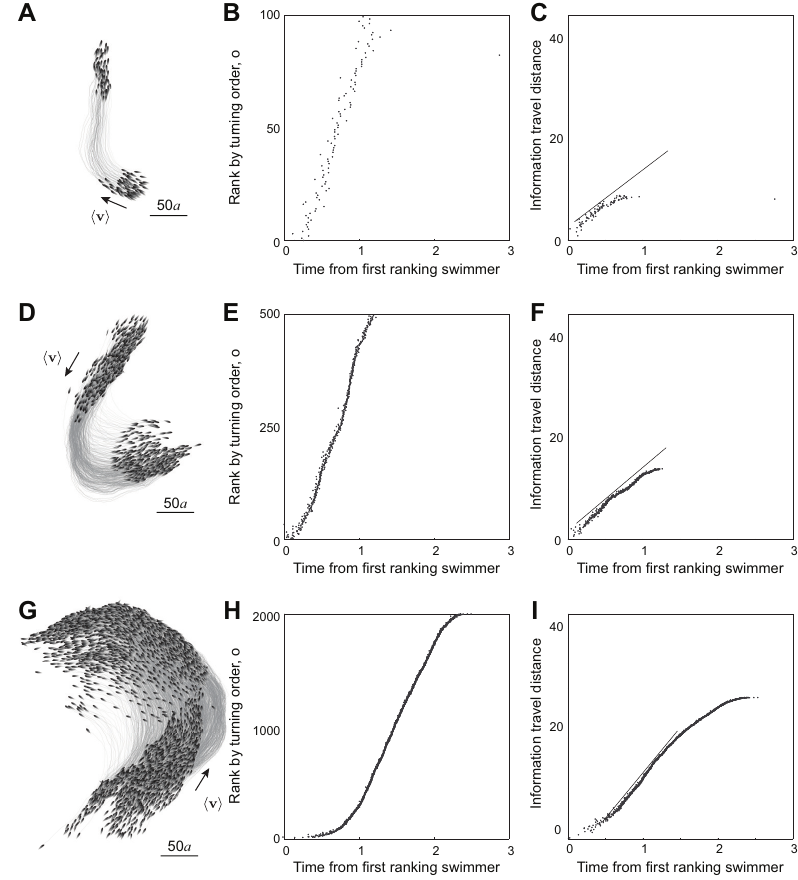} 
\caption[]{\footnotesize \textbf{Analysis of additional turning events.} 
\textbf{A.}  Turning trajectories of a school containing $100$ fish and \textbf{B.} corresponding rank of fish by the order $o$ at which they reach maximal curvature and \textbf{C.} information travel distance defined as $\sqrt{o/\textrm{density}}$ versus absolute turning time delay. The information transfer speed is 11.4 times the individual swimming speed $U$. 
\textbf{D.}  Turning trajectories of a school containing $500$ fish and \textbf{E.} corresponding rank of fish by the order $o$ at which they reach maximal curvature and \textbf{F.} information travel distance defined as $\sqrt{o/\textrm{density}}$ versus absolute turning time delay. The information transfer speed is 12.7 times the individual swimming speed $U$. 
\textbf{G.}  Turning trajectories of a school containing $2000$ fish and \textbf{H.} corresponding rank of fish by the order $o$ at which they reach maximal curvature and \textbf{I.} information travel distance defined as $\sqrt{o/\textrm{density}}$ versus absolute turning time delay. The information transfer speed is 18.2 times the individual swimming speed $U$. 
}
\label{figsi:turn}
\end{figure*}

\begin{figure*}[!t]
\centering
\includegraphics[scale =1]{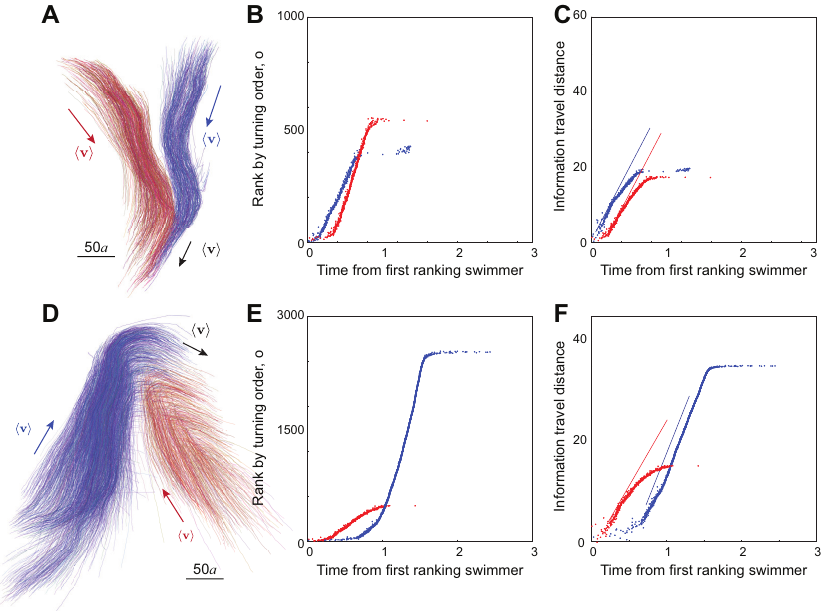}
\caption[]{\footnotesize \textbf{Analysis of additional merging events.}
\textbf{A.}  Turning trajectories of a school containing $1000$ fish and \textbf{B.} corresponding rank of fish by the order $o$ at which they reach maximal curvature and \textbf{C.} information travel distance defined as $\sqrt{o/\textrm{density}}$ versus absolute turning time delay. The information transfer speeds of both clusters are 28.7 times the individual swimming speed $U$. 
\textbf{D.}  Turning trajectories of a school containing $3000$ fish and \textbf{E.} corresponding rank of fish by the order $o$ at which they reach maximal curvature and \textbf{F.} information travel distance defined as $\sqrt{o/\textrm{density}}$ versus absolute turning time delay. The information transfer speeds of both clusters are  26.0 and 37.8 times the individual swimming speeds $U$ for the red and blue subgroups, respectively.
}
\label{figsi:merge}
\end{figure*}

\begin{figure*}[!t]
\centering
\includegraphics[scale =1]{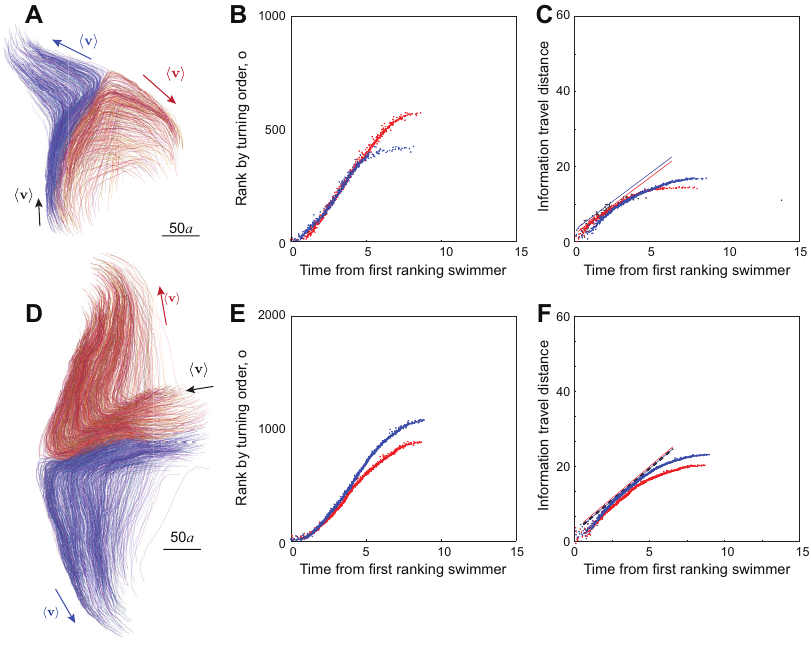}
\caption[]{\footnotesize \textbf{Analysis of additional splitting events.}
\textbf{A.}  Turning trajectories of a school containing $1000$ fish and \textbf{B.} corresponding rank of fish by the order $o$ at which they reach maximal curvature and \textbf{C.} information travel distance defined as $\sqrt{o/\textrm{density}}$ versus absolute turning time delay. The information transfer speeds of both clusters are 3.0 times the individual swimming speed $U$. 
\textbf{D.}  Turning trajectories of a school containing $2000$ fish and \textbf{E.} corresponding rank of fish by the order $o$ at which they reach maximal curvature and \textbf{F.} information travel distance defined as $\sqrt{o/\textrm{density}}$ versus absolute turning time delay. The information transfer speeds of both clusters are 3.4 times the individual swimming speed $U$. 
}
\label{figsi:split}
\end{figure*}

\begin{figure*}[!t]
\centering
\includegraphics[scale =1]{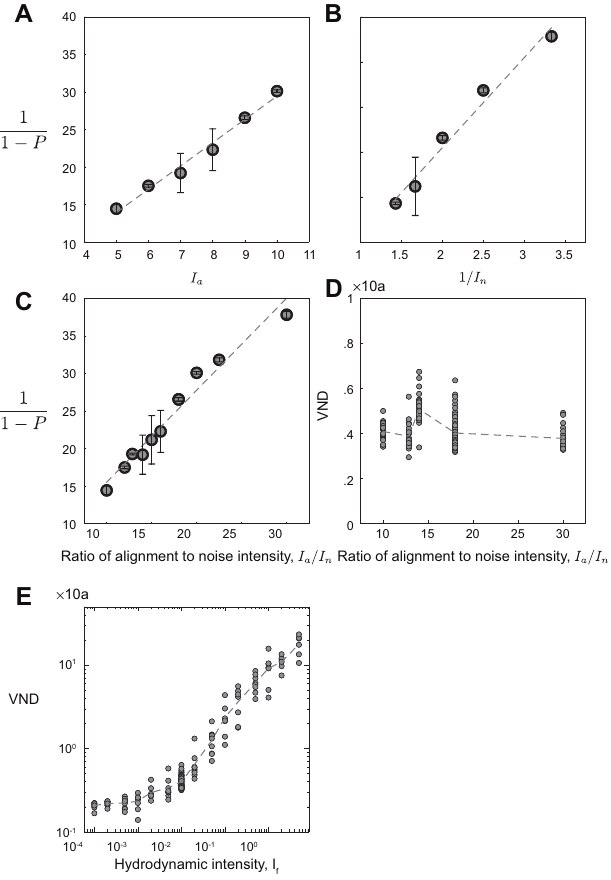}
\caption[]{\footnotesize \textbf{Scaling  with alignment, noise and hydrodynamic intensities.}
\textbf{A.} Polar order parameter $1/ (1-P)$ plotted as a function of alignment intensity $I_a$. Noise intensity is kept at fixed value $I_n=0.5$. The fitting curve is $1/ (1-P) = 3.11 I_a +-1.61$ with $R^2=0.98$. 
\textbf{B.} Polar order parameter $1/(1-P)$ plotted versus the inverse of noise intensity $1/ I_n$. The fitting curve is $1/ (1-P) = 9.99 / I_n + 5.49$ with $R^2=0.98$. 
Alignment intensity is kept at fixed value $I_a=9$. 
\textbf{C.} Polar order parameter $1/ (1-P)$ plotted as a function of ratio between alignment intensity and noise intensity $I_a/ I_n$ including all simulations from panels {A} and {B}.
The fitting curve is $1/ (1-P) = 1.23 I_a / I_n + 3.23$ with $R^2=0.964$. 
\textbf{D.} Average distance to Voronoi neighbors as a function of $I_a/ I_n$. 
In A-D, hydrodynamic  $I_f=0.01$, $N=$100 to 10,000, 
$(I_a, I_n) = (9,0.5), \, (9,0.3), \, (9,0.7), \, (5,0.5), \, (7,0.5)$. 
\textbf{E.} Average distance to Voronoi neighbors as a function of hydrodynamic intensity $I_f$. Parameter values: $I_a=9$, $I_n=0.5$. 
In all panels, five Monte Carlo simulations are performed for each parameter set, each for a total integration time of $T = 1000$. 
}
\label{figsi:IaIn}
\end{figure*}

\singlespacing
\begin{table}[!h]
\centering
\caption{\textbf{Summary of the dataset generated numerically. } 
We performed and analyzed 631 distinct simulations at various parameter values and school sizes, each for a total integration time $T = 1000$. 
}
\label{tab:summary}
\begin{tabular}{cccccccc}
        \toprule
        $I_a$ & $I_n$ & $I_f$ & $N$ & $\Delta N$& \#MC & \# & $P$\\ 
        \midrule
       9 & 0.5 & 0.01 &100  & - & 5 & 5& 0.96 \\
       9 & 0.5 & 0.01 & 1000  & -& 5 & 5 & 0.79 \\
       9 & 0.5 & 0.01 & 10,000 & - & 5 & 5 &  0.69 \\
       9 & 0.5 & 0.01 & 50,000  & -& 1 & 1 &  0.81 \\
       \midrule
       9 & 0.5 & 0.01 & 110-540  & 10 & 1 & 44 & 0.87-0.96 \\
       9 & 0.5 & 0.01 & 550-900  & 50  & 1& 7 & 0.78-0.89 \\
       9 & 0.5 & 0.01 &1,500  & -& 1& 1 & 0.78 \\
       9 & 0.5 & 0.01 &1,600  & -& 1& 1 & 0.68 \\
       9 & 0.5 & 0.01 &2,000  & -& 5& 5 & 0.83 \\
       9 & 0.5 & 0.01 &2,500  & -& 1& 1 & 0.76 \\
       9 & 0.5 & 0.01 &3,000  & -& 7& 7 & 0.73 \\
       9 & 0.5 & 0.01 &3,600  & -& 1& 1 & 0.74 \\
       9 & 0.5 & 0.01 &4,900  & -& 1& 1 & 0.80 \\
       9 & 0.5 & 0.01 &5,000  & -& 7& 7 & 0.67 \\
       9 & 0.5 & 0.01 &6,400 & - & 1& 1 & 0.59 \\
       9 & 0.5 & 0.01 &7,500  & -& 6& 6 & 0.77 \\
       9 & 0.5 & 0.01 &8,100  & -& 1& 1 & 0.73 \\
        \midrule
        9 & 0.5 & $10^{-4}$ -- 5  &100, 200, 500, 1000 & - & 5 & 375& 0.67-0.98 \\
        9 & 0.5 & $10^{-4}$ -- 5  &1500, 2000, 2500, 3000  & -& 1 & 60 & 0.66-0.98 \\
        9 & 0.5 & $10^{-4}$ -- 5 &10,000  & -& 1 & 15& 0.26-0.95  \\
         \midrule
        9 & 0.5 & 0 &100, 1000, 10,000  & - & 10 & 30 & 0.96-0.99 \\
         \midrule
       5 & 0.5 & 0.01 &100-1000 & 100& 1 & 10& 0.83-0.92 \\   
       5 & 0.5 & 0.01 &5000 & -& 1 & 1&   \\ 
       7 & 0.5 & 0.01 &100-1000 & 100& 1 & 10& 0.87-0.95 \\
       9 & 0.7 & 0.01 &100-1000 & 100 & 1 & 10& 0.80-0.94 \\
       9 & 0.3 & 0.01 &100-1000 & 100& 1 & 10& 0.9-0.97 \\  
        9 & 0.3 & 0.01 &5000 & -& 1 & 1&   \\  
         \midrule
       9 & 0.0 & 0.01 &100, 1000, 10,000  & - & 1 &3 & 0.92-0.98 \\
       9 & 0.75 & 0.01 &100, 1000, 10,000  & - & 1 &3& 0.73-0.87 \\
       9 & 1.0 & 0.01 &100, 1000, 10,000  & - & 1 &3& 0.63-0.70 \\
       \bottomrule
\end{tabular}
\end{table}


\end{document}